\documentstyle[12pt]{article}
\def\nn{\nonumber\\}

\input epsf
\begin{document}
\title{Quantum Algorithm, Gaussian Sums, \\and Topological Invariants}
\author{
K. Shiokawa\thanks {E-mail address: kshiok@phys.sinica.edu.tw}
\\
{\small Institute of Physics, Academia Sinica,} \\
{\small Nankang, Taipei 11529, Taiwan}}
\date{\today}
\maketitle
\begin{abstract}
Certain quantum topological invariants of three manifolds can be written in the form of the Gaussian sum.
It is shown that such topological invariants can be approximated
efficiently by a quantum computer.
The invariants discussed here are obtained as a partition
function of the gauge theory on three manifolds with various gauge groups.
Our algorithms are applicable to Abelian
and finite gauge groups and to some classes of non-Abelian gauge groups. These invariants can be directly estimated by the nuclear magnetic resonance (NMR)
technique used for evaluating the Gaussian sum.
\end{abstract}
\maketitle

\newpage

\section{Introduction and Summary}

Despite of the expectation for potentially wide applications of quantum computation\cite{NielsenChuang00},
the speedup due to quantum algorithms is limited only to restricted problems.
Well-known examples with exponential improvement such as the factorization and discrete-log algorithm\cite{Shor97} both depend on the quantum Fourier transformation as a crucial technique. The similar method was later extended to other number theoretic algorithms\cite{Hallgren01,DamSeroussi02}.

Experimental implementations of these algorithms are even more limited
due to the difficulty of maintaining coherence during
quantum computation even for a small number of qubits.
Among them, the most remarkable example is a realization of the
Shor's factorization algorithm\cite{VSBYS01} by the NMR technique. The factorization of
larger numbers was also demonstrated recently by the NMR
technique\cite{MMAMS07} through evaluating the Gaussian sum.
These methods crucially rely on the experimental determination of
periodicity in a function.
It is worth investigating to find more applications based
on similar techniques.

In this work, we explore different applications of the Gaussian sum through its relation to
the topological invariants of
three dimensional manifolds.
The evaluation of topological invariants of three manifolds has been
discussed in relations with topological quantum
computation\cite{FKLW02,FKW02}. A similar line of thoughts lead to
efficient algorithms for the estimation of topological invariants.
The efficient algorithm approximating Jones
polynomial was constructed\cite{AJL05,WocjanYard06,GMR07}.
 Although these approaches are theoretically elegant and intriguing, experimental
realization of these algorithms is still far from a practical stage.

Quantum topological invariants can be obtained essentially as partition
functions of some special classes of quantum fields defined on three manifolds\cite{Witten88}.
The further simplification may be possible to reduce the partition function to the
form with its periodicity manifest. In some cases, it can be reduced
to the Gaussian sum form.

In this paper, we discuss three cases in which this reduction is
possible. Each case is related to an interesting physical theory.
In Sec. 2.1, we discuss Abelian gauge theory invariants. They are defined as
partition functions of the Abelian Chern-Simons gauge field on the underlying three
manifold. A large class of such invariants can be written as
Gaussian sums. In Sec. 2.2, we discuss non-Abelian gauge theory
invariants, defined as partition functions of the non-Abelian Chern-Simons gauge field
on three manifolds. We study the case in which these variants can be reduced to Gaussian sums..
Finally in Sec. 2.3, we discuss the case in which the gauge group is finite. In
this case, we obtain the invariants known as Dijkgraaf-Witten invariants\cite{DijkgraafWitten90}.
In these cases, we show that efficient approximation algorithms
based on the evaluation of multivariate Gaussian sums exist.
These invariants can be experimentally observed by the NMR technique for factorizing large numbers and by other optical processes.

\section{Wilson loop as a topological invariant}

The Chern-Simons action for the $SU(2)$ gauge field $A$ is
 \begin{eqnarray}
 S[A]=\frac{k}{4\pi}\int_{M} d^3 x
 \epsilon^{ijk} {\mbox Tr} \left[ A_i \partial_j A_k
  +\frac{2i}{3} A_i A_j A_k
 \right].
       \label{LagrangianforANA}
\end{eqnarray}
The partition function we study is
 \begin{eqnarray}
Z[M]=\int_{M} DA e^{iS[A]}.
       \label{PartitionFn}
\end{eqnarray}
 The observable we study is the Wilson loop $\langle W(C) \rangle$
defined along an oriented curve $C$ as
 \begin{eqnarray}
 W(C)=\mbox{Tr}_R \left[ P e^{i e \oint_{C}  A_{\mu}dx^{\mu}} \right],
       \label{W}
       \end{eqnarray}
where the trace is taken in the irreducible representation of a
gauge group. By closing a loop and taking a trace, $W(C)$ is defined
to be gauge-invariant.
When the curve $C$ is a link $L$ consisting of disjoint knot
components $C_1,...,C_m$,
 \begin{eqnarray}
W(L)= W(C_1,...,C_m)=\prod_{j=1}^{m}  W(C_j).
       \label{WCm}
       \end{eqnarray}
The expectation value of the Wilson loop is given by
 \begin{eqnarray}
\langle W(L) \rangle=\int_{M} DA ~W(L) e^{iS[A]}.
       \label{Wilsonloopexp}
\end{eqnarray}

Since $\langle W(L) \rangle$ does not involve any metric, we expect
that it is essentially topologically invariant.
In order to see the exact topological invariance,
we consider the Dehn surgery representation of the three manifold $M$
with a framed link $L$ embedded in $S^3$ as follows:
First consider a torus obtained as a tubular neighborhood of each
component of $L$. We denote a union of all $m$ tubular neighborhoods as
$N(L)$. We can decompose ${S^3}$ as a union between ${S^3}-$int $N(L)$
and $m$ tori. The Dehn surgery is defined by pasting each torus back
to where it was in ${S^3}$ with a twist such that a meridian of
each torus and each longitude of ${S^3}-$int $N(L)$ is identified.
Arbitrary closed oriented three manifold $M$ can be obtained
by Dehn surgeries\cite{Lickorish62} around a link $L$ embedded in ${S^3}$. Thus
regarding a twist as a framing of each component of a link,
topological information of $M$ can be obtained by studying a framed
link in ${S^3}$.

Two three-manifolds obtained by different framed links are
homeomorphic to each other if and only if they are related by Kirby moves\cite{Kirby78}.
In particular, any topological invariant needs to to be invariant under Kirby moves.
Kirby moves consist of adding or removing a trivial knot $C$ with its self-linking number $\pm 1$
and making a connected sum of a component of a link with another along their framing.
It turns out that the expectation value $\langle W(L) \rangle$ is not invariant
under Kirby moves but aquires $+1$,
a phase
as
\begin{eqnarray}
 \langle W(L) \rangle \rightarrow e^{\pm i \theta} \langle W(L) \rangle,
         \label{WKirby}
       \end{eqnarray}
       where $\theta=3\pi(k-2)/4k $ (assuming $k\geq 2$)
is related to the central charge in two dimensional theory.
Thus $\langle W(L) \rangle$ itself is not a topological invariant
and needs some modification.
We will come back to this in Sec. 2.2.

\subsection{Abelian gauge group}

We construct the topological invariant for the Abelian gauge field
$A$ in the following. For the Abelian gauge field $A$ defined on the three manifold $M$,
the second term in (\ref{LagrangianforANA}) vanishes and its action $S$ becomes
 \begin{eqnarray}
 S[A]=\frac{k_0}{8\pi}\int_{M} d^3 x
 \epsilon^{ijk} A_i \partial_j A_k.
       \label{LagrangianforA}
\end{eqnarray}

We consider a link composed of $m$ components $C_1,...,C_m$ embedded
in  $M=S^3$.
The vacuum expectation value of the Wilson loop is related to the
correlation function for the gauge field as
\begin{eqnarray}
\langle W(C_1,...,C_m) \rangle&=& \langle \prod_{j=1}^{m} e^{i e_j
\oint_{C_j} A_{\mu}dx^{\mu}} \rangle\nn &=& 1 - \sum_{i,j=1}^{m}
e_i e_j \oint_{C_i} \oint_{C_j}\langle A_{\mu}(x_i) A_{\nu}(x_j)
\rangle dx_i^{\mu}dx_j^{\nu} +\cdots \nn &=& \exp\left[ -
\sum_{i,j=1}^{m} e_i e_j \oint_{C_i} \oint_{C_j}\langle
A_{\mu}(x_i) A_{\nu}(x_j) \rangle dx_i^{\mu}dx_j^{\nu} \right] \nn
      \label{C1Cm}
\end{eqnarray}
The correlation function of the Wilson loop can be written as
\begin{eqnarray}
\langle A_{\mu}(x_j) A_{\nu}(x_l) \rangle = \frac{i}{k_0} \int_{C_j}
dx^j \int_{C_l} dy^l \epsilon^{jlk} \frac{(x-y)^k}{|x-y|^3}.
      \label{AA}
\end{eqnarray}
If we write the charges $e_i$ as integer multiples of an elementary
charge $e$ as $e_i=n_i e$ ($n_i=1,2,...$) and, by absorbing $e$ in the coupling constant as $k\equiv k_0/e^2$, then the correlation
function has the following form\cite{Polyakov88}:
 \begin{eqnarray}
\langle  W(C_1,...,C_m) \rangle=
e^{-i\frac{2\pi}{k}\left(\sum_j n_j^2 r_{j}+ \sum_{j \neq l} n_j n_l
J_{jl} \right)},
      \label{Wilsonmloop}
\end{eqnarray}
where
\begin{eqnarray}
J_{ij}=\frac{1}{2\pi} \int_{C_i} dx^i \int_{C_j} dy^j \epsilon^{ijk}
\frac{(x-y)^k}{|x-y|^3}
      \label{J}
\end{eqnarray}
is a linking number of links $C_i$ and $C_j$ and $r_i$ is a
self-linking number. Note that $J_{ij}$ is a symmetric integer
matrix and $r_i$ is an integer vector. Naive definition of $r_i$
\begin{eqnarray}
r_{i}=\frac{1}{2\pi} \int_{C_i} dx^i \int_{C_i} dy^j \epsilon^{ijk}
\frac{(x-y)^k}{|x-y|^3}
      \label{r}
\end{eqnarray}
contains divergence and we need to regularize it. In order not to
sacrifice the metric independence by explicitly introducing the cutoff parameter, we choose the point splitting regularization prescription by using framing of links, i.e. deforming the contour $C$ parametrized as
$x^{\mu}(s)$ to the new contour $C_{\delta}$ by
$x^{\mu}(s)\rightarrow x^{\mu}(s)+\delta x^{\mu}(s)$ and define
\begin{eqnarray}
r_{\delta i}=\lim_{\delta x \rightarrow 0} \frac{1}{2\pi} \int_{C}
dx^i \int_{C_{\delta}} dy^j \epsilon^{ijk} \frac{(x-y)^k}{|x-y|^3},
      \label{ri}
\end{eqnarray}
which is a self-linking term between the original knot and its framing.

The Wilson loop is invariant under the shift of variables $n_i
\rightarrow n_i + k$ for an integer $k$. Hereafter we only consider
this case. This will limit the range of $n_i$ to be finite:
$n_i=1,2,...,k$, where each $n_i$ gives a different representation of $U(1)$.
  In order to obtain the quantity independent of the representation, we sum over all representations
  and write this sum as $\langle W(L) \rangle$ as follows
 \begin{eqnarray}
\langle W(L) \rangle=\sum_{n_i,n_j=1}^{k} e^{-i\frac{2\pi}{k}\left(\sum n_i^2 r_{\delta i}+ \sum_{i\neq j} n_i n_j J_{ij} \right)}.
      \label{WilsonL}
\end{eqnarray}
If we write $J_{ii}=r_{\delta i}$ in above,
\begin{eqnarray}
 \langle W(L) \rangle=\sum_{n_i,n_j=1}^{k} e^{-i\frac{2\pi}{k}\left(\sum_{i,j=1}^{m} n_i n_j J_{ij} \right)}.
 \label{IMnoselfterm}
\end{eqnarray}
This can be viewed as a partition function for the spin variables
$n_i$ that take $k$ different values at the vertices $i$ of the graph.
The shape of the graph is determined by the edges connecting vertices
$i$ and $j$ for $J_{ij}\neq 0$.

We assume that $k=p^m$ with an odd prime $p$,
$J_{ij}$ can be
diagonalized modulo $k$ with a matrix $U\in SL(m,Z)$\cite{MOO92}. Let us
denote the components after diagonalization as $J_1, ..., J_m$.
Then $U^{T} J U= \bigoplus_{i=1}^k J_i$ and
\begin{eqnarray}
 \langle W(L) \rangle=\prod_{i=1}^{m} \left( \sum_{n_i} e^{-i\frac{2\pi}{k}n_i^2 J_{i}
 } \right).
 \label{AbelianWilsonLGauss}
\end{eqnarray}
For $k\equiv 1~(\mbox{mod}~ 4)$, $\langle W(L) \rangle$ itself can be shown to be a topological
invariant\cite{Guadagnini93}. In this case,
the Abelian invariant $ \tau_A(M)$ can be written as a product of Gauss sums:
\begin{eqnarray}
 \tau_A(M)=\prod_{i=1}^{m} G(k,J_i),
 \label{IMGauss}
 \end{eqnarray}
where
\begin{eqnarray}
 G(k,a)=
 \sum_{n=0}^{k-1} e^{-i\frac{2\pi}{k} a n^2}.
 \label{Gauss1}
\end{eqnarray}

Appearance of $G(k,a)$ indicates that $\tau_A(M)$ is  computationally hard to evaluate as the estimation of the Gaussian sum is considered to be classically hard. On the other hand, the use of quantum algorithms  enables us to approximate $\tau_A(M)$ in polynomial time, as each Gaussian sum factor can be done so\cite{DamSeroussi02}.

 While the norm of the sum is straightforward to calculate, its phase requires $O(1/\epsilon)$
times measurements to estimate within the error $\epsilon$.
The Gauss sum is defined by
\begin{eqnarray}
 G(k,a)=
 \sum_{n=0}^{k-1} \chi(n) e^{-i\frac{2\pi}{k} a n }
 \label{GaussDef}
\end{eqnarray}
$\chi(n)$ is a multiplicative character. For our purpose, we take
$\chi(n)$ to be the Legendre symbol:
\begin{eqnarray}
 \chi(n)=\left( \frac{n}{k} \right). \label{Legendre}
\end{eqnarray}
It is convenient to define $\chi(0)=0$.
For an odd prime $k$,
\begin{eqnarray}
 \left( \frac{n}{k}\right)=
        n^{(k-1)/2} (\mbox{mod} ~k) ~\mbox{for} ~n\neq 0
\label{Legendreexp}
\end{eqnarray}
holds.
One can easily show that $G(k,a)$ for an integer $a$ is equivalent
to the familiar form given in (\ref{Gauss1})\cite{BEW98}.

The Gauss sum can be efficiently estimated by the following steps\cite{DamSeroussi02}.
First we construct a quantum state with the coefficient given by the Legendre symbol.
Second we use a unitary transformation to this state to change the coefficient to the Gauss sum.
Then we make a measurement to obtain the amplitude of the state closely related to this state
that gives the estimation of the Gauss sum.

Preparation of a state:
\begin{eqnarray}
 | \chi \rangle=\frac{1}{\sqrt{k-1}}
 \sum_{n=0}^{k-1} \chi(n) | n \rangle
 \label{chistate}
\end{eqnarray}
can be done in the following way.
We use the fact that the following quantum Fourier transform
\begin{eqnarray}
 | p \rangle \longrightarrow \frac{1}{\sqrt{k}}
 \sum_{s=0}^{k-1} e^{-i\frac{2\pi}{k} p s } | s \rangle
 \label{QFT}
\end{eqnarray}
can be performed efficiently on a quantum
computer by O($k^2$) steps\cite{NielsenChuang00}.
Making the quantum Fourier transformation on the second term on the
product state $|n\rangle|1\rangle$ gives
\begin{eqnarray}
|n\rangle \otimes \frac{1}{\sqrt{k}}\sum_{l=0}^{k-1}q^{-l} |l\rangle,
 \label{nFT}
\end{eqnarray}
where $q\equiv e^{i\frac{2\pi}{k}}$.
By shifting each $|l\rangle$ by $\frac{k-1}{2} \log_q n$
\begin{eqnarray}
&|n\rangle& \otimes \frac{1}{\sqrt{k}}\sum_{l=0}^{k-1}q^{-l} |l +\frac{k-1}{2} \log_q n\rangle\\ \nonumber
=&|n\rangle& \otimes \frac{1}{\sqrt{k}}\sum_{l=0}^{k-1}q^{-l+\frac{k-1}{2} \log_q n} |l\rangle
\\ \nonumber
=&q^{\frac{k-1}{2} \log_q n}&|n\rangle \otimes \frac{1}{\sqrt{k}}\sum_{l=0}^{k-1}q^{-l} |l\rangle
\\ \nonumber
=&\chi(n)&|n\rangle \otimes \frac{1}{\sqrt{k}}\sum_{l=0}^{k-1}q^{-l} |l\rangle,
 \label{nFTashift}
\end{eqnarray}
where the base $q$ of the logarithm is viewed as the element in $Z_k$. $\log_q n$
is an integer $p$ such that $q^p=n (\mbox{mod} k)$.
The evaluation of the discrete logarithm $\log_q n$ can be efficiently performed on a quantum computer
by poly($\log_2 k$) steps\cite{Shor97}.

We make another quantum Fourier transform and write the right hand side in (\ref{chistate}) as
\begin{eqnarray}
 | \chi \rangle&=&\frac{1}{\sqrt{k-1}}\frac{1}{\sqrt{k}}
 \sum_{n=0}^{k-1} \chi(n) \sum_{l=0}^{k-1} e^{-i\frac{2\pi}{k} a n l } | l
 \rangle \nn
 &=&\frac{1}{\sqrt{k-1}}\frac{1}{\sqrt{k}}
 \sum_{l=1}^{k-1} \chi(l^{-1}) \sum_{n=0}^{k-1} \chi(n)  e^{-i\frac{2\pi}{k} a n} | l
 \rangle \nn
 &=&\frac{1}{\sqrt{k-1}}\frac{G(k,a)}{\sqrt{k}}
 \sum_{l=1}^{k-1}  \chi(l^{-1})| l
 \rangle, \label{chistateQFT}
\end{eqnarray}
where we used
\begin{eqnarray}
 G(k,al)=
 \left\{ \begin{array}{cc}
     \chi(l^{-1}) G(k,a) & \mbox{for} ~l\neq 0\\
        0      &\mbox{for} ~l=0
         \end{array}      \right.
\label{GayGy}
\end{eqnarray}
from the first line to the second line. Mapping $| l
 \rangle$ to $\chi(l)^2 | l
 \rangle$ by following the same procedure as in (\ref{nFTashift}) will make the last term in ({\ref{chistateQFT}) into
\begin{eqnarray}
\frac{G(k,a)}{\sqrt{k}}
 | \chi \rangle.
 \label{chistateQFT2}
\end{eqnarray}
Knowing that $|G(k,a)|=\sqrt{k}$\cite{BEW98}, now the phase information for
the Gauss sum is stored in the coefficient.

By operating a phase gate which adds the phase
$G(k,a)/\sqrt{k}=e^{-i\phi}$ on the state $|\chi \rangle$ by the above
procedure and does nothing on the state $|0\rangle$, we
can transform the initial state $|0\rangle+|\chi \rangle$ to
\begin{eqnarray}
 |0\rangle+|\chi
\rangle&\leftarrow&|0\rangle+e^{-i\phi}|\chi \rangle.
 \label{zerochistate}
\end{eqnarray}
Then by making observation in the order of $1/\epsilon$ times, we can
determine the phase $\phi$ within the error $\epsilon$.

Using the algorithm explained above, evaluating each Gaussian sum
appeared in our topological invariant
\begin{eqnarray}
 \tau_A(M)=\prod_{i=1}^{m} G(k,J_i)
 \label{IMGauss2}
 \end{eqnarray}
takes poly($k$,$1/\epsilon$) time. Thus the evaluation of the invariant can be
done in $m$ $\times$ poly($k$,$1/\epsilon$) = poly($m,k,1/\epsilon$) time.

\subsection{Non-Abelian gauge group}

$\langle W(L) \rangle$  can be considered naturally as a link polynomial.
For $M=S^3$, it is possible to choose the framing such that
all self-linking terms vanish (called standard framing).
In this case,
$\langle W(L) \rangle$ only depends on the link $L$ and becomes
proportional to the Jones polynomial $J(L,q)$.
$\langle W(L) \rangle$ can be calculated explicitly by
the combinatorial method similar to the one for the Jones polynomial.
\begin{figure}[h]
 \begin{center}
\epsfxsize=.45\textwidth \epsfbox{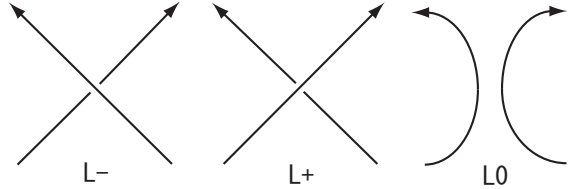}
\end{center}
\caption{ The skein relation\label{fig1}}
\end{figure}

By assigning the same color $n=2$ (corresponding to two-dimensional irreducible representation of $sl_2(C)$) for all link components, above
 $\langle W(L) \rangle$ with local deformations shown in Fig. 1 obeys the
 following skein relation
\begin{eqnarray}
q^{1/4}\langle W(L_{+})\rangle-q^{-1/4}\langle W(L_{-})\rangle=(q^{1/2}-q^{-1/2})\langle W(L_{0})\rangle.
         \label{skeinW}
       \end{eqnarray}
$W(L)$ still depends on the framing. In order to cancel frame
dependence, we define $\langle \tilde{W}(L)\rangle$ by multiplying $\langle W(L)\rangle$ with
$[2]^{-1}e^{-3\pi i w(L)/2k}$,
where $[n]\equiv \sin\left(n\pi/k\right)/\sin\left(\pi/k\right)$ and $w(L)$ is a writhe defined by a difference of the number of positive
and negative crossings.
Then we see that $\langle\tilde{W}(L)\rangle$ satisfies
\begin{eqnarray}
q\langle\tilde{W}(L_{+})\rangle-q^{-1}\langle\tilde{W}(L_{-})\rangle
=(q^{1/2}-q^{-1/2})\langle\tilde{W}(L_{0})\rangle.
         \label{skeinW2}
       \end{eqnarray}
Comparing this with the skein relation satisfied by the Jones
polynomial $J(L)$
\begin{eqnarray}
t^{-1}J(L_{+})-t J(L_{-})=(t^{1/2}-t^{-1/2})J(L_{0}),
         \label{skeinJ}
       \end{eqnarray}
       we see $\langle \tilde{W}(L)\rangle$ for $n=2$
       is essentially the Jones
polynomial $J(L_{+})$ by identifying $t^{1/2}=-q^{-1/2}$.
Thus we define $\tilde{J}(L)$, a version of a Jones polynomial proportional to $\langle \tilde{W}(L)\rangle$
by normalizing $\tilde{J}(L)$ as $\tilde{J}(U)=1$ for a trivial knot $U$.
The explicit relation between these is obtained in \cite{KirbyMelvin91} as
\begin{eqnarray}
\langle\tilde{W}(L)\rangle =[2] e^{3\pi i/2k \sum_{i,j=1}^{m} J_{ij} } \tilde{J}(L).
         \label{WfromJall}
       \end{eqnarray}

Mathematically rigorous definition of the invariants is given
in \cite{ReshetikhinTuraev90}.
It can be expressed in our formula (\ref{IM}) as follows.
For each link component $C_i$ whose framing $r_i$,
we sum over all the color $n_i$ with proper weights.
By writing the color for each link and the sum explicitly in (\ref{IM}), we have
\begin{eqnarray}
\tau_{NA}(M)= c^m e^{-i\theta\sigma_L}\sum_{n_i}[n_1]...[n_m]\langle W_{n_1...n_m}(L)
\rangle,
         \label{tauMsumn}
       \end{eqnarray}
where $c\equiv (2/k)^{1/2}\sin\left(\pi/k\right)$.
Here we followed the definition in \cite{KirbyMelvin91}, which is different from
that in \cite{ReshetikhinTuraev90} by an overall factor $c^{\nu}$, where $\nu$
is the first Betti number of $M$.

The evaluating $\tau_{NA}(M)$ for general $k$ is difficult.
Neverthess, for certain values of $k$, the calculation can
be simplified. In particular, for $k=3$, $[2]=1$ and $\tilde{J}(L)=1$ in (\ref{WfromJall}) gives
\begin{eqnarray}
\langle\tilde{W}(L)\rangle =e^{\pi i \sum_{i,j=1}^{m} J_{ij}/2 }.
         \label{WfromJk3}
       \end{eqnarray}
In \cite{KirbyMelvin91}, this result is used to write the invariant as
\begin{eqnarray}
\tau_{NA}(M)= 2^{-m/2} e^{-i\pi
\sigma_L/4}\sum_{S\subset L}e^{i\pi \sum_{i,j\in S} J_{ij}/2},
         \label{tauMr3KM}
       \end{eqnarray}
       where the sum is over all the sublink $S$ with color $n=2$.
       It is easy to see that this can also be written as follows:
\begin{eqnarray}
\tau_{NA}(M)= 2^{-m/2} e^{-i\pi
\sigma_L/4}\sum_{n_i=1}^{2}e^{i\pi \sum_{i,j} J_{ij}n_i n_j/2}.
         \label{tauMr3Gauss}
       \end{eqnarray}
Thus it is reduced to the multivariate Gaussian sum.
From our result in Sec. 2.1, this sum can be estimated
in $m$ $\times$ poly($1/\epsilon$). Combining with the calculation $\sigma_L$, that takes typically $O(m^2)$ time,  we see that $\tau_{NA}(M)$ can be estimated in poly($m,1/\epsilon$) time.

\subsection{Finite gauge group}
Geometrically the gauge field can be viewed as a connection on a
principal G bundle over $M$. Then the path integral is a sum over
all connections on $M\times G$ with the gauge group $G$. For a compact simply-connected
gauge group, a principal G bundle is topologically trivial and the path integral
over connections is reduced to the ordinary path integral of the gauge field.
In our case, the action is given by the Chern-Simons action in (\ref{LagrangianforANA}).

For a finite gauge group $G$, a principal G bundle has a unique flat
connection and nontrivial contribution to the path integral is
coming only from a sum over the paths with different topologies. In other words,
the path integral is replaced by the discrete sum over different
conjugate classes of the G-bundle. Each conjugate class of the
G-bundle is represented by the holonomy $\gamma$ for the map $\pi_1(M)\rightarrow G$. In \cite{DijkgraafWitten90}, the partition function for the gauge field with a finite gauge group is considered.
The path integral is given by the sum over all homotopy classes of $\gamma$ as
\begin{eqnarray}
\tau_{DW}(M)= \frac{1}{|G|} \sum_{\gamma} e^{2\pi i S_{\gamma} },
         \label{DW}
       \end{eqnarray}
where $e^{2\pi i S_{\gamma}}=\langle \gamma^{*}\alpha,[M]\rangle$
       is a pairing between the three-dimensional homology class of $M$, $[M]\in H_3(M,U(1))$
       and the cohomology class $\gamma^{*}\alpha$
       for each $\alpha\in H^3(BG;U(1))$ pulled back by regarding
       $\gamma$ as a map from $M$ to the classifying space $BG$.
       For a flat connection, the
       principle $G$ bundle is completely specified by the homotopy
       class of $\gamma$.

For $G=Z_k$, more explicit expression can be obtained\cite{MOO92}.
For the action $S_{\gamma}$ to be real,  $\langle \gamma^{*}\alpha,[M]\rangle\in
U(1)$. Since $H^3(BZ_k;U(1))\cong Z_k$, choosing $\alpha$ is equivalent to choosing an integer $l \in {1,...,k}$ and $e^{2\pi i l/k} \in U(1)$.
Meanwhile $\gamma$ has a corresponding element $\gamma_0\in H^1(M;Z_k)$ by using Hom$(\pi_1(M),Z_k) \simeq $Hom$(H_1(M;Z),Z_k)\simeq H^1(M;Z_k)$.
Let $\delta^{*}$ be a connecting homomorphism between $H^1(M;Z_k)$ to $H^2(M;Z)$
associated with an exact sequence $0\rightarrow Z\stackrel{k}{\rightarrow}
Z\rightarrow Z_k\rightarrow 0$.
Then $\gamma_0 \cup \delta^{*}\gamma_0$ defines an element in $H^3(M;Z_k)$.
Exponentiating this will give an element in $H^3(M,U(1))$.
Coupling this element with $[M]$ gives an action $e^{2\pi i S_{\gamma}}=\langle \gamma^{*}\alpha,[M]\rangle = \langle e^{2\pi i l (\gamma_0 \cup \delta^{*}\gamma_0)/k} ,[M] \rangle$.
$\gamma_0 \cup \delta^{*}\gamma_0$ can be evaluated as the intersection number among dual chains
and given as $S_{\gamma}=\sum_{i,j=1}^{m} J_{ij}n_i n_j/k$\cite{MOO92}, where the linking number $J_{ij}$ appears when $M$ is represented by Dehn surgeries along the link. Thus we obtain
\begin{eqnarray}
\tau_{DW}(M)= \frac{1}{|G|} \sum_{n_i=1}^{k-1} e^{2\pi i \sum_{i,j=1}^{m}J_{ij}n_i
n_j/k},
         \label{DWGauss}
       \end{eqnarray}
       where we took $l=k$.
       Thus again the invariant $\tau_{DW}(M)$ can be written as the multivariate Gaussian sum, thus can be estimated efficiently with the method developed in Sec. 2.1.



\end{document}